\documentclass[a4paper,11pt]{article}
\pdfoutput=1

\usepackage[T1]{fontenc}
\usepackage{amsmath,amssymb,amsfonts}
\usepackage{graphicx}
\usepackage{hyperref}
\usepackage[numbers,sort&compress]{natbib}
\usepackage{authblk}

\hypersetup{
  colorlinks=true,
  linkcolor=blue,
  citecolor=blue,
  urlcolor=blue
}

\title{BBN constraints on primordial black holes with a continuous memory-burden crossover}

\author[1,2]{Xuan-Yu Zhang}
\author[1,3]{Mei-Ting Yang}
\author[1,4,5]{Hong-Bo Jin\thanks{Corresponding author: \href{mailto:hbjin@bao.ac.cn}{hbjin@bao.ac.cn}}}

\affil[1]{National Astronomical Observatories, Chinese Academy of Sciences, Beijing 100101, China}
\affil[2]{School of Physics and Optical Engineering, Zhejiang University of Technology, Hangzhou 310014, China}
\affil[3]{School of Physics and Astronomy, Yunnan University, Kunming 650500, China}
\affil[4]{School of Astronomy and Space Science, University of Chinese Academy of Sciences, Beijing 100049, China}
\affil[5]{The International Center for Theoretical Physics Asia-Pacific (ICTP-AP), University of Chinese Academy of Sciences, Beijing 100190, China}

\date{}

\begin{document}
\maketitle

\begin{abstract}
Light primordial black holes (PBHs) are disfavored as dark matter if they evaporate through standard Hawking radiation alone. The memory-burden effect can extend their lifetimes by suppressing emission after roughly half the mass is lost. Existing cosmological studies often model the onset of this phase as an instantaneous transition between semi-classical and burden-dominated evaporation. We instead treat the crossover as continuous and compare additive versus multiplicative combinations of the two rates, using a smoothed tanh profile with parameters $(q,\delta)$. Monochromatic PBHs are mapped to a decaying scalar field and evolved with Modified AlterBBN during Big Bang nucleosynthesis (BBN). The two prescriptions yield distinct exclusion curves: the additive crossover always gives weaker bounds than the multiplicative one, while both are tighter than the instantaneous transition. For $10^{5}\,\mathrm{g}\lesssim M_i\lesssim 10^{10}\,\mathrm{g}$, the additive case can permit $f_{\mathrm{PBH},0}\sim 10^{-1}$ where the multiplicative case gives $f_{\mathrm{PBH},0}\lesssim 10^{-2}$. Specifying the rate-combination rule is therefore essential when translating memory-burden models into BBN constraints on PBH dark matter.
\end{abstract}

\medskip
\noindent\textbf{Keywords:} primordial black holes; Big Bang nucleosynthesis; particle physics--cosmology connection; quantum black holes

\section{Introduction}
\label{sec:intro}

Primordial black holes (PBHs) may form in the early Universe and remain among the best-motivated non-particle dark matter candidates \cite{carr2021constraints,green2021primordial}. Their observational fate is tied to mass loss through Hawking radiation \cite{hawking1975particle,hawking1974black}. For a Schwarzschild black hole, the semi-classical evaporation rate is \cite{hawking1975particle,montefalcone2025does}
\begin{equation}
\left. \frac{d M}{d t} \right|_{SC} = - \frac{\mathcal{G} g_{\star, H} (T_{BH}) M_{Pl}^{4}}{30720 \pi M^{2}} \label{eq:SC}
\end{equation}
where $\mathcal{G} \approx 3.8$ is a greybody factor \cite{page1976particle}, $M_{Pl} \approx 2.2 \times 10^{-5} \mathrm{~g}$ is the Planck mass, and $g_{\star, H}$ counts relativistic degrees of freedom with $T < T_{BH}$ \cite{macgibbon1990quark,macgibbon1991quark}. Lower-mass black holes are hotter and evaporate faster. Equation~\eqref{eq:SC} implies that PBHs with $M \lesssim 5 \times 10^{14} \mathrm{~g}$ disappear within the age of the Universe \cite{montefalcone2025does,carr2021constraints}. Current limits then disfavor PBHs with $M \lesssim 10^{17} \mathrm{~g}$ as the dominant dark matter component \cite{carr2021constraints,green2021primordial}.

Cosmological probes of evaporating PBHs are especially sensitive during Big Bang nucleosynthesis (BBN). If a significant PBH population evaporates between $t \sim 0.1\,\mathrm{s}$ and $t \sim 10^{4}\,\mathrm{s}$, injected entropy and high-energy particles can alter light-element yields \cite{cyburt2016big,carr2010new,keith2020constraints,acharya2020cmbbbn}. BBN therefore provides a sharp test of any scenario in which light PBHs survive longer than in the standard evaporation picture. Below we examine how the onset of memory burden modifies the evaporation history probed by BBN.

Recent quantum-gravity-motivated work suggests that part of the excluded mass window may reopen \cite{dvali2020black,dvali2024memory,alexandre2024new,thoss2024breakdown}. In Dvali's quantum $N$ portrait \cite{dvali2013black,dvali2014black}, a black hole is a condensate of $N$ soft gravitons with nearly gapless ``memory modes'' that encode holographic information \cite{dvali2018microscopic,dvali2019universe}. Hawking emission lowers the master-mode occupation number and pushes the system away from a critical maximal-packing state. The resulting memory burden resists further decay \cite{dvali2018microscopic,dvali2024memory}: stored patterns cost energy to maintain, and evaporation slows once roughly half the mass is lost \cite{dvali2020black,dvali2019universe}. A prototype description uses master, memory, and decay sectors \cite{dvali2020black}
\begin{equation}
\hat{H} = \epsilon_{0} \hat{n}_{0} + \sum_{k} \epsilon_{k} \left(1 - \frac{\hat{n}_{0}}{N_{c}}\right)^{p} \hat{n}_{k} + \epsilon_{0} \hat{m}_{0} + C_{0} \left(\hat{a}_{0}^{\dagger} \hat{b}_{0} + \hat{b}_{0}^{\dagger} \hat{a}_{0}\right) \label{eq:H}
\end{equation}
where $\hat{n}_{0}$, $\hat{n}_{k}$, and $\hat{m}_{0}$ count master, memory, and external quanta, $N_{c}$ is the critical master occupation, and $\hat{b}_{0}^{\dagger}\hat{a}_{0}$ describes Hawking emission. As $\hat{n}_{0}$ drops, the burden parameter $\mu = \sum_k n_k \partial \mathcal{E}_k / \partial n_0$ rises and suppresses further emission \cite{dvali2019universe}. Light PBHs can then survive to today; for example, $M_i \sim 10^{4}\,\mathrm{g}$ may survive for about a Hubble time \cite{dvali2020black,dvali2024memory,alexandre2024new}.

To make contact with cosmology, previous studies often compress this physics into a two-stage evaporation law \cite{alexandre2024new,thoss2024breakdown,montefalcone2025does}. Stage~I follows Eq.~\eqref{eq:SC} until $M = q M_i$; Stage~II follows a burden-suppressed rate,
\begin{equation}
\frac{d M}{d t} = 
\begin{cases} 
\left. \frac{d M}{d t} \right|_{SC} & \text{if } M > q M_i \\ 
\left. \frac{d M}{d t} \right|_{MB} & \text{if } M < q M_i 
\end{cases}
\label{eq:piecewise}
\end{equation}
with
\begin{equation}
\left. \frac{\mathrm{d} M}{\mathrm{d} t} \right|_{\mathrm{MB}} = \frac{1}{\tilde{S}(q M_i)^{k}} \cdot \left. \frac{\mathrm{d} M}{\mathrm{d} t} (q M_i) \right|_{\mathrm{SC}} \label{eq:MB}
\end{equation}
where $\tilde{S}$ is the dimensionless Bekenstein entropy \cite{bekenstein1973black}, $k$ sets its scaling, and we take $k=2$ as in Refs.~\cite{montefalcone2025does,dvali2024memory,alexandre2024new,thoss2024breakdown}. This piecewise form is useful, but it turns on the memory burden instantly at $M = q M_i$. Microscopic arguments and recent phenomenology instead suggest a gradual crossover \cite{alexandre2024new,montefalcone2025does}.

In this paper we quantify the dependence of BBN limits on the implementation of the crossover between semi-classical and burden-dominated evaporation. Ref.~\cite{montefalcone2025does} studied a smooth crossover with only the multiplicative rate combination. We extend that analysis by (i)~comparing additive and multiplicative combinations on equal footing and (ii)~quantifying the resulting shift in BBN exclusion curves. Section~\ref{sec:pheno} introduces the crossover models. Section~\ref{sec:bbn} describes the scalar-field mapping and AlterBBN setup, presents the constraints, and interprets the differences. Section~\ref{sec:conclusion} summarizes our findings.

\section{Crossover models for memory-burdened evaporation}
\label{sec:pheno}

We start from the instantaneous prescription in Eqs.~\eqref{eq:piecewise}--\eqref{eq:MB} and replace the sharp transition with a smooth function of mass. No exact mass-evolution solution is known once memory burden builds up continuously, so this step is explicitly phenomenological. The goal is not to derive $h(M)$ from Eq.~\eqref{eq:H}, but to quantify how observational limits shift when the crossover is smoothed and the branch rates are combined in different ways.

The piecewise law switches at $M = q M_i$. A Heaviside step function $H(x)$ encodes this behavior:
\begin{equation}
H(x) = 
\begin{cases} 
0 & \text{if } x < 0 \\ 
\frac{1}{2} & \text{if } x = 0 \\ 
1 & \text{if } x > 0 
\end{cases}
\label{eq:step}
\end{equation}
Following Ref.~\cite{montefalcone2025does}, we set $x = (M - q M_i)/(\delta q M_i)$ and write $H(M) = H(x)$, so that $H=0$ for $M < q M_i$ and $H=1$ for $M > q M_i$. The dimensionless parameter $\delta$ controls the crossover width.

Two prescriptions for combining the semi-classical and burden-dominated rates are available. The additive prescription is
\begin{equation}
\frac{d M}{d t} = H(M) \left. \frac{d M}{d t} \right|_{SC} + [1 - H(M)] \left. \frac{d M}{d t} \right|_{MB} \label{eq:add}
\end{equation}
and the multiplicative prescription is
\begin{equation}
\frac{d M}{d t} = \left( \left. \frac{d M}{d t} \right|_{SC} \right)^{H(M)} \left( \left. \frac{d M}{d t} \right|_{MB} \right)^{1 - H(M)} \label{eq:mul}
\end{equation}
Ref.~\cite{montefalcone2025does} used only Eq.~\eqref{eq:mul}. With the exact step function the two forms give the same $\mathrm{d}M/\mathrm{d}t$. Once the step function is replaced by a smooth profile $h(M)$, they need not coincide because the branch rates differ by many orders of magnitude near the crossover.

For a continuous crossover we take $h(M)$ to be monotonic and to approach $H(M)$ as $\delta \to 0$. Ref.~\cite{montefalcone2025does} also discusses error-function, arctangent, and rational profiles; for numerical stability and direct comparison, we use
\begin{equation}
h_{1}(M) = \frac{1}{2} \left[ 1 + \tanh\left(\frac{M - q M_i}{\delta q M_i / 2}\right) \right] \label{eq:tanh}
\end{equation}
inserted into Eqs.~\eqref{eq:add} and \eqref{eq:mul} in place of $H(M)$. The same $(q,\delta)$ and the same $h_1(M)$ are used for both prescriptions, so any difference in the BBN curves arises solely from additive versus multiplicative blending.

\section{BBN constraints}
\label{sec:bbn}

\subsection{Observational setup}
\label{sec:bbn-setup}

Standard BBN successfully predicts primordial light-element abundances once the Universe cools through $T \sim 3\,\mathrm{MeV}$ to $T \sim 10\,\mathrm{keV}$ \cite{cyburt2016big,fields2020big}. Any energy injection in that interval can disturb the neutron-to-proton ratio, dissociate nuclei, or alter the effective baryon density \cite{kawasaki2005big,poulin2016fresh,pospelov2010big,carr2010new,keith2020constraints}. PBHs with masses low enough to evaporate appreciably during BBN are therefore strongly constrained, regardless of whether evaporation follows the standard Hawking law or a modified memory-burden history.

Memory burden lengthens PBH lifetimes and can keep lighter objects active during BBN \cite{dvali2020black,dvali2024memory,alexandre2024new}. Ref.~\cite{montefalcone2025does} computed BBN limits for a smooth multiplicative crossover and found that moderate $(q,\delta)$ exclude PBHs with $M_i \lesssim 4 \times 10^{16}\,\mathrm{g}$ from comprising all of dark matter. Here we use the same monochromatic setup and AlterBBN framework and directly compare additive and multiplicative crossovers.

\subsection{Scalar-field mapping and numerical method}
\label{sec:bbn-method}

AlterBBN is designed to evolve decaying species during BBN \cite{arbey2012alterbbn,arbey2020alterbbn}. To use it for PBHs, we map a monochromatic population to an equivalent decaying scalar field, as is standard for evaporating relics \cite{kawasaki2005big,poulin2016fresh,carr2010new}. A scalar field obeys
\begin{equation}
\dot{\rho}_{\phi} = - 3 H \rho_{\phi} - \Gamma_{\phi} \rho_{\phi} \label{eq:scalar}
\end{equation}
while non-relativistic PBHs with $\rho_{\mathrm{PBH}} = n_{\mathrm{PBH}} M_{\mathrm{PBH}}$ evolve as
\begin{equation}
\dot{\rho}_{\mathrm{PBH}} = - 3 H \rho_{\mathrm{PBH}} - \left(- \frac{\dot{M}_{\mathrm{PBH}}}{M_{\mathrm{PBH}}}\right) \rho_{\mathrm{PBH}} \label{eq:pbh}
\end{equation}
The Hubble dilution term is identical in both cases. The evaporation term maps onto scalar decay once we define the effective rate $\Gamma_{\mathrm{eff}} = -\dot{M}_{\mathrm{PBH}}/M_{\mathrm{PBH}}$ and update it at each time step from the crossover model of Section~\ref{sec:pheno}. Evaporation products cascade to long-lived photons and hadrons to which BBN is sensitive \cite{macgibbon1991quark,carr2010new,keith2020constraints}, so the two descriptions probe the same observables.

\begin{figure}[tbp]
    \centering
    \includegraphics[width=\textwidth]{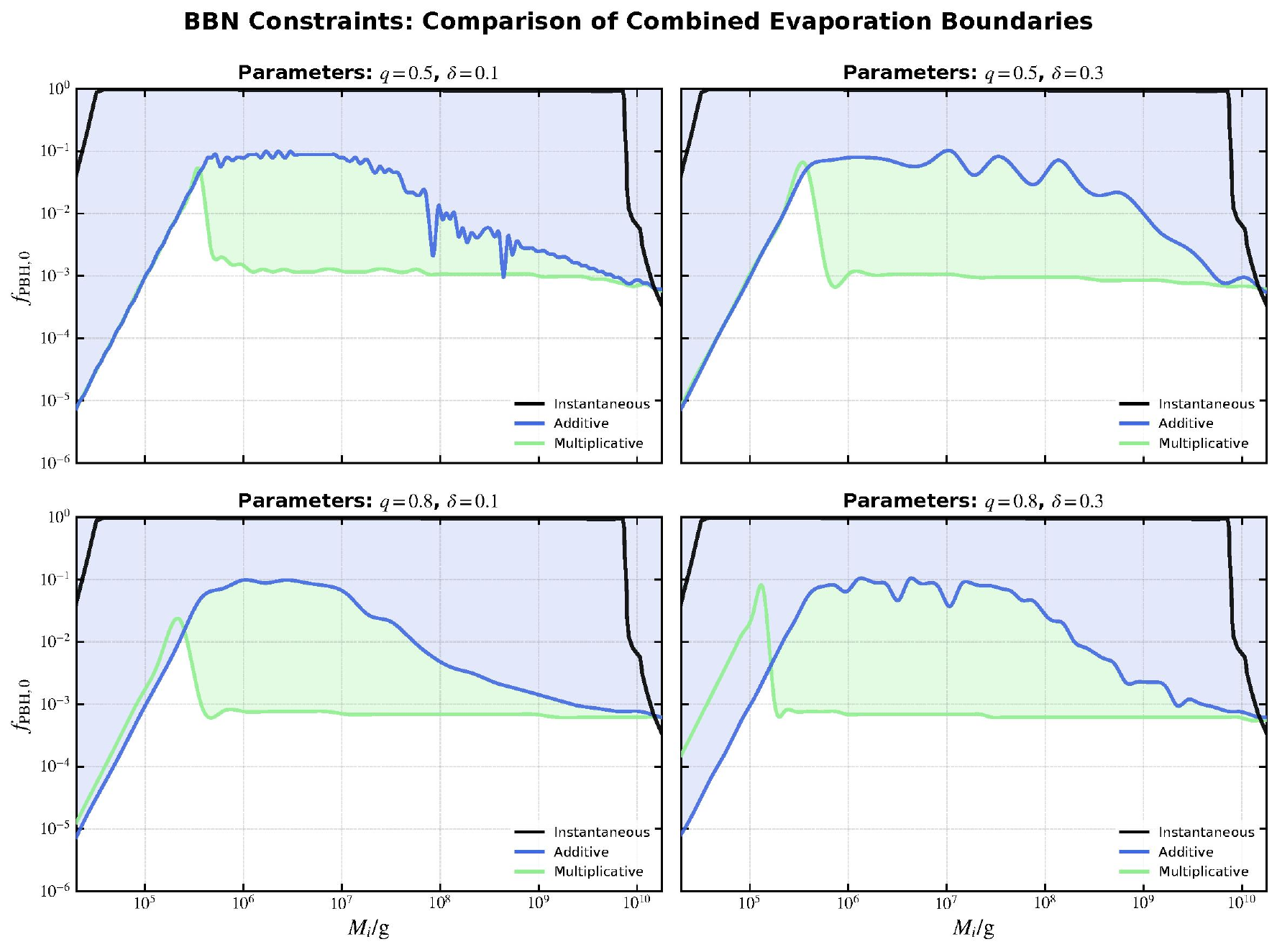}
    \caption{BBN upper limits on the initial PBH fraction $f_{\mathrm{PBH},0}$ as a function of initial mass $M_i$ on log--log scales. Four panels correspond to $(q,\delta)=(0.5,0.1)$, $(0.5,0.3)$, $(0.8,0.1)$, and $(0.8,0.3)$, arranged from left to right and top to bottom. Black curves: instantaneous transition \cite{montefalcone2025does}. Blue curves: additive crossover. Green curves: multiplicative crossover. Shaded regions above each curve are excluded; the green band marks the separation between the additive and multiplicative bounds. For $10^{5}\,\mathrm{g}\lesssim M_i\lesssim 10^{10}\,\mathrm{g}$, the multiplicative bound is tighter than the additive one.}
    \label{fig:bbn_constraints}
\end{figure}

We implement $\Gamma_{\mathrm{eff}}$ from Eqs.~\eqref{eq:add}--\eqref{eq:tanh} in Modified AlterBBN \cite{montefalcone2025does,arbey2020alterbbn} and scan the initial PBH fraction $f_{\mathrm{PBH},0}$ at fixed $M_i$. The largest $f_{\mathrm{PBH},0}$ consistent with light-element data defines the exclusion boundary in fig.~\ref{fig:bbn_constraints}. We show four representative choices $(q,\delta) = (0.5,0.1)$, $(0.5,0.3)$, $(0.8,0.1)$, and $(0.8,0.3)$, together with the limit from the instantaneous transition in Ref.~\cite{montefalcone2025does}.

To validate the numerical setup, we adopt the same monochromatic mapping, tanh crossover profile, greybody factor, and light-element inputs as Ref.~\cite{montefalcone2025does}. For the multiplicative prescription, our implementation follows that reference; the additive curves are new to this work.

\subsection{Results and interpretation}
\label{sec:bbn-results}

Figure~\ref{fig:bbn_constraints} summarizes the BBN bounds on the maximum allowed initial fraction $f_{\mathrm{PBH},0}$ at fixed $M_i$. Black curves show the instantaneous transition \cite{montefalcone2025does}. Blue and green curves show additive and multiplicative smooth crossovers, respectively. In each panel, any point above a curve is excluded at that $M_i$. The multiplicative bound lies below the additive one throughout $10^{5}\,\mathrm{g} \lesssim M_i \lesssim 10^{10}\,\mathrm{g}$, so it is the tighter constraint in that range. Both smooth bounds sit below the instantaneous curve over the same range, meaning both crossovers are more restrictive than the instantaneous transition, with the additive case always the weaker of the two.

The spread between prescriptions is largest where PBHs evaporate mainly during BBN, roughly $10^{5}\,\mathrm{g} \lesssim M_i \lesssim 10^{10}\,\mathrm{g}$ \cite{montefalcone2025does,thoss2024breakdown}. In that window the additive crossover can permit $f_{\mathrm{PBH},0} \sim 10^{-1}$ while the multiplicative curve typically stays at $f_{\mathrm{PBH},0} \lesssim 10^{-2}$, for the same $(q,\delta)$. Physically, the additive form keeps a larger semi-classical contribution until $M$ falls below $q M_i$, so the total injection rate stays higher for longer. Because the two branch rates differ by orders of magnitude, additive blending resembles the arithmetic mean of the branch rates and multiplicative blending the geometric mean. The additive history is therefore closer to the instantaneous transition, yet still constrained by the smooth crossover.

These results have a direct cosmological implication. When memory burden is modeled with a continuous crossover, BBN limits quoted from a single rate combination can over- or underestimate the allowed PBH fraction. A conservative exclusion should use the multiplicative prescription, while the additive case defines the systematic spread associated with the combination rule.

\section{Conclusion}
\label{sec:conclusion}

We studied BBN constraints on monochromatic PBHs whose evaporation crosses smoothly from a semi-classical phase to a memory-burden-dominated phase. Using the tanh crossover profile $h_1(M)$ and Modified AlterBBN, we compared additive and multiplicative combinations of the two evaporation rates.

The crossover prescription matters. The additive model always gives weaker BBN bounds than the multiplicative one for the parameters explored, while both are tighter than the instantaneous transition. For $10^{5}\,\mathrm{g} \lesssim M_i \lesssim 10^{10}\,\mathrm{g}$, the difference can reach order unity in $f_{\mathrm{PBH},0}$. Any future cosmological analysis of memory-burdened PBHs should therefore state the chosen combination rule alongside $(q,\delta)$.

Several extensions are natural within the same framework: other smooth crossover functions $h(M)$, extended mass distributions, and complementary probes such as CMB anisotropies \cite{acharya2020cmbbbn,montefalcone2025does} or indirect-detection limits \cite{carr2021constraints,green2021primordial}. The present BBN study already shows that phenomenological crossover modeling is not a minor detail for light PBH dark matter.

\section*{Data availability}
\label{sec:data}

The BBN exclusion boundaries in fig.~\ref{fig:bbn_constraints} are the primary numerical output of this work; no separate data release accompanies the manuscript. The monochromatic limits were computed with AlterBBN \cite{arbey2012alterbbn,arbey2020alterbbn}, implementing the Modified AlterBBN setup of Ref.~\cite{montefalcone2025does} to evolve a decaying scalar field with the time-dependent rate $\Gamma_{\mathrm{eff}}$ defined in Section~\ref{sec:pheno}. Reproduction requires the public AlterBBN release together with the modified-evaporation prescription given in Ref.~\cite{montefalcone2025does}.

\section*{Acknowledgments}

This work is funded by the National Astronomical Observatories of the Chinese Academy of Sciences, Project No.~E4TG6601, and has been supported in part by the National Key Research and Development Program of China under Grant No.~2021YFC2203000.

\bibliographystyle{JHEP}
\bibliography{References}

\end{document}